\documentclass[aps,prl,showpacs,twocolumn,superscriptaddress,longbibliography]{revtex4-1}
\usepackage[pdftex]{graphicx}
\usepackage[pdftex,bookmarks,colorlinks,breaklinks]{hyperref} 
\hypersetup{linkcolor=red,citecolor=blue,filecolor=dullmagenta,urlcolor=blue} 
\newcommand{\eqref}[1]{(\ref{#1})}
\begin{document}

\title{Waveguide photonic limiters based on topologically protected resonant modes}
\author{U. Kuhl}
\affiliation{Universit\'e de Nice-Sophia Antipolis, Laboratoire de la Physique de la Mati\`ere Condens\'ee, CNRS, Parc Valrose, 06108 Nice, France}
\author{F. Mortessagne}
\affiliation{Universit\'e de Nice-Sophia Antipolis, Laboratoire de la Physique de la Mati\`ere Condens\'ee, CNRS, Parc Valrose, 06108 Nice, France}
\author{E. Makri}
\affiliation{Department of Physics, Wesleyan University, Middletown CT-06459, USA}
\author{I. Vitebskiy}
\affiliation{Air Force Research Laboratory, Sensors Directorate, Wright-Patterson Air Force Base, OH-45433, USA}
\author{T. Kottos}
\affiliation{Department of Physics, Wesleyan University, Middletown CT-06459, USA}


\begin{abstract}
We propose a concept of chiral photonic limiters utilising topologically protected localised midgap defect states in a photonic waveguide. The chiral symmetry alleviates the effects of structural imperfections and guaranties a high level of resonant transmission for low intensity radiation. At high intensity, the light-induced absorption can suppress the localised modes, along with the resonant transmission. In this case the entire photonic structure becomes highly reflective within a broad frequency range, thus increasing dramatically the damage threshold of the limiter. Here we demonstrate experimentally the principle of operation of such photonic structures using a waveguide consisting of coupled dielectric microwave resonators.
\end{abstract}

\maketitle

Photonic limiters are protecting filters transmitting input signal with low power (or energy) while blocking the signals of excessively high power (or energy) \cite{1,2,3,4,5,6,7}. Usually, a passive limiter absorbs the high-level signal, which can cause its overheating or other irreversible damage. The input level above which the transmitted signal intensity doesn't grow with the input is the limiting threshold (LT). Another critically important characteristic is the limiter damage threshold (LDT), above which the limiter sustains irreversible damage. The domain between LT and LDT is the dynamical range of the limiter - the larger it is, the better. Unfortunately, material limitations impose severe restrictions on both thresholds. It is, therefore, imperative to utilise appropriate photonic platforms which are both flexible enough to provide simultaneously tunable and low LT and high enough LDT. Importantly, these structures should provide broadband protection and they should be tolerant to deviation of the material and geometrical parameters from their ideal values.
Along these lines, the defect modes hosted by photonic band-gap structures have been exploited as an alternative to achieve flexible, high efficiency photonic limiters. In most occasions, however, limiting action is achieved by a non-linear frequency shift of the transparency window of the photonic structure \cite{8,9,10,11,12,13,14}. Such a shift is inherently small and, therefore, cannot provide broadband protection from high-power input. To address this issue we have recently proposed a concept of a reflective photonic limiter \cite{15,16}. Such a limiter does not absorb but reflects the high radiation, thereby, protecting itself - not just the receiving device. The principle of operation is based on resonant transmission via a localised defect mode with purely nonlinear absorption. The resonant frequency is engineered to fall in the middle of a photonic band-gap, allowing for the structure to provide a broadband protection in case of self-destruction of the defect mode. This self-destruction occurs at high input intensity, which triggers nonlinear losses suppressing the localised mode. As a result, the incident harmful signal experiences nearly total reflection. In contrast, at low intensity the losses at the nonlinear defect are negligible and a strong resonant transmission will take place.

\begin{figure*}
  \centerline{\includegraphics[width=0.8\linewidth]{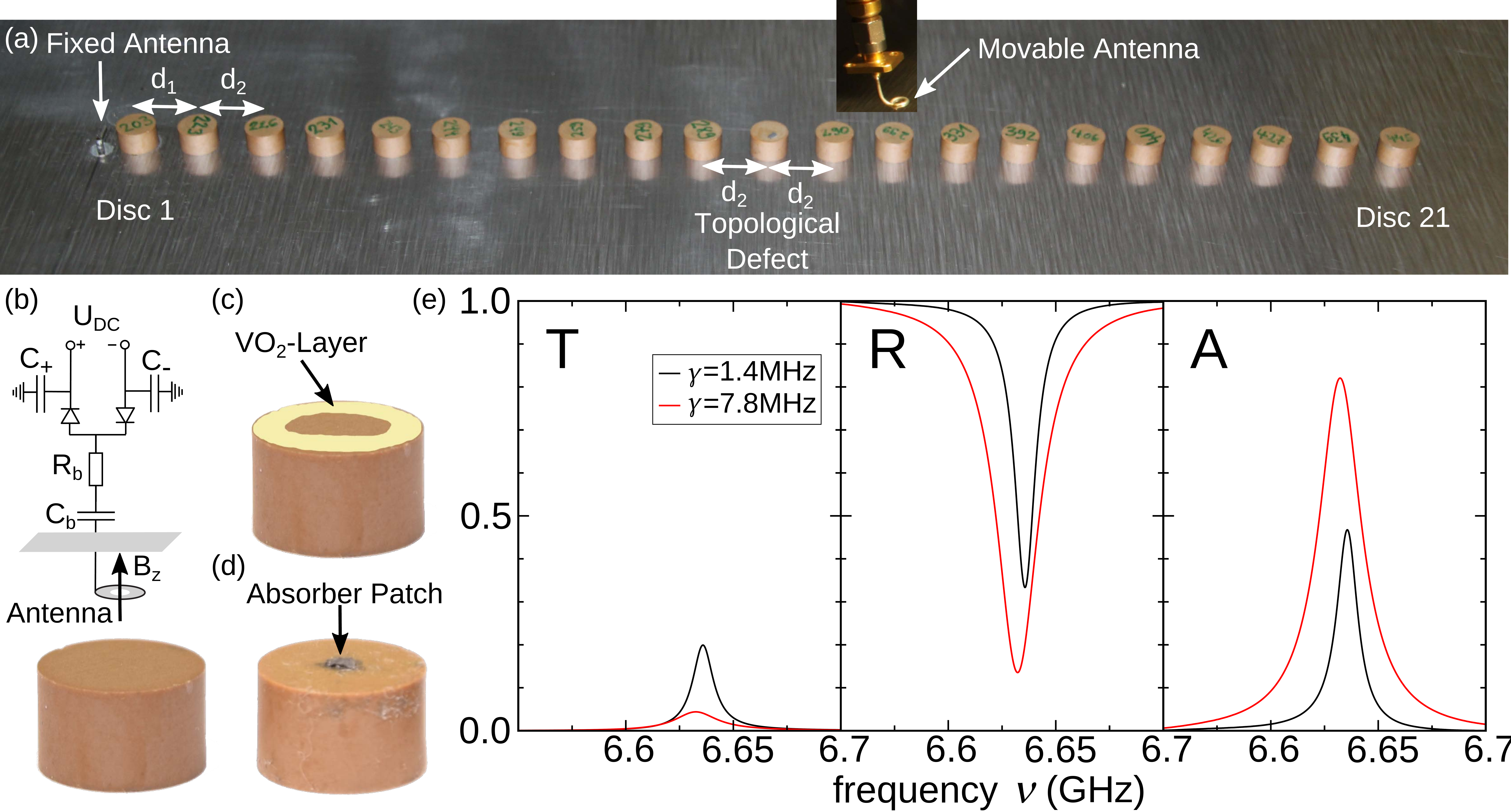}}
  \caption{\label{fig:fig1} {\bf Experimental set-up and single resonator properties.} (a) The experimental set-up consists of 21 identical coupled dielectric cylindrical resonators sandwiched between two metallic plates. The resonators are separated by distances $d_1$ or $d_2$ with $d_1<d_2$. A central dimerisation defect is introduced by repeating the spacing $d_2$. The defect creates a topological interface state at frequency equal to the eigenfrequency of the isolated resonators. (b) A proposal for the implementation of non-linear losses in the defect resonator. It involves various module (sensing antenna, diode, threshold DC voltage). (c) An alternative proposal for the defect resonator involves an epitaxial growth of a thin film consisting of a material that experiences a thermally induced insulator-to-metal phase transition. (d) Our measurements involve a defect resonator, which includes a manually modulated absorbing patch. (e) Measured transmittance $T$, reflectance $R$ and absorption $A$ for two different patches. The linewidth $\gamma$ (1.4 and 7.8 MHz) of the reflected signal is mainly characterises the losses due to the absorbing patches. The observed small shift to the resonance frequency from the value of $\nu_0$=6.6545\,GHz to lower values is due to the coupling of the resonator to the kink antenna.
  }
\end{figure*}

Here we propose the use of coupled resonator waveguide arrays (CROW) \cite{17,18,19,20} with alternating short and long distances from one another (see Fig.~\ref{fig:fig1}), as a fertile platform to implement structurally robust reflective waveguide limiters with a wide dynamic range. This CROW design possesses a chiral (sub-lattice) symmetry. In the presence of a defect, this symmetry provides topological protection to a midgap defect mode \cite{21,22,23,24}. The defect state is strongly localised around the defect resonator and has staggering form with non-vanishing components at alternating resonators \cite{22,23}. For low incident power (or energy) it can provide high transmittance shielded from (positional) fabrication imperfections. When (non-linear) losses at the defect resonator (triggered due to heating from high power - or energy- incident radiation or other non-linear lossy mechanisms) exceed a critical value, the resonant defect mode and the associated resonant transmission are dramatically suppressed. The whole photonic structure becomes highly reflective and not absorptive for a broad frequency range. As a result, the LDT increases with a consequent increase of the dynamic range of the limiter. We demonstrate experimentally this operational principle of absorption-enhanced reflectivity using a microwave CROW arrangement with chiral symmetry, thus paving the way to a new class of limiters with wide dynamic range and a reconfigurable LDT.

\section*{Results}
\subsection{Experimental set-up.}

The set-up consists of a chain of $N=21$ high index cylindrical resonators (radius r=4mm, height h=5mm, made of ceramics with an index of refraction $n\approx 6$) with eigenfrequency around $\nu_0=6.655$\,GHz and linewidth $\gamma=1.4$\,MHz. The resonators are placed at alternatively distances $d_1=12$\,mm and $d_2=14$\,mm corresponding to strong ($t_1=38$\,MHz) and weak ($t_2=21$\,MHz) evanescent couplings, respectively. A topological defect at the 11th resonator is introduced by repeating the spacing $d_2$. On the left hand side of the array, close to the first resonator, we have placed a kink antenna that emits a signal. The structure is shielded from above with a metallic plate (not shown) where a movable loop antenna (receiving antenna) is mounted and it is coupled to the 13th resonator. A photograph of the experimental set-up is shown in Fig.~\ref{fig:fig1}(a).

Next we assume that the defect resonator incorporates a nonlinear absorption mechanism, i.e.\ we assume that its losses are self-regulated depending on the strength of the incident field radiation. One option to incorporate nonlinear losses is via an external element (fast diodes), see Fig.~\ref{fig:fig1}(b). It provides "on-the-fly" re-configurability of the LT via an externally tuned DC voltage $U_\textrm{DC}$. The magnetic field, due to the transmitted signal, at the defect resonator is coupled to the diodes via a loop antenna. Its strength dictates the value of the induced current at the antenna and consequently the voltage across the diodes.  The latter dictates if the diodes are "on" (high voltage corresponding to high losses at the defect resonator) or "off" (low voltage corresponding to low losses). An alternative nonlinear lossy mechanism is associated with temperature driven insulator-to-metal phase transition materials like Vanadium dioxide (VO$_2$) \cite{25,26,27,28}, which can be deposited on top of the defect resonator [see Fig.~\ref{fig:fig1}(c)].

In our experiment we are not concerned with the physical origin of the nonlinear losses at the defect resonator. Rather we focus on demonstrating their effects on the transport properties of the photonic structure and thus establishing the operational principle of (structurally) robust reflective photonic limiters with wide dynamical range. Therefore, we have included losses by placing an absorbing patch on top of the resonator [see Fig.~\ref{fig:fig1}(d)]. This process results in a slight shift of the real part of the permittivity of the defect resonator, which we corrected by using resonators with slightly higher eigenfrequency. The linewidth $\gamma$ has been used in order to quantify the losses, of the resonators. In Fig.~\ref{fig:fig1}(e) we show the transmittance $T$, reflectance $R$, and absorption $A=1-T-R$ for two resonators with different losses. The transmittance $T$, is measured from the kink antenna to the loop antenna, which was positioned above the resonator. The reflectance $R$, is measured from the kink antenna. We observe that the stand-alone lossy resonator reduces the transmittance $T$ as the losses increases, thus acting as a limiter. However this reduction in $T$ comes to the expense of increasing absorption, i.e.\ the stand-alone lossy resonator acts as a sacrificial limiter.

\subsection{Mathematical model.} The photonic structure is described by a one-dimensional (1D) tight-binding Hamiltonian
\begin{equation}\label{eq:eq1}
H_P=\sum_n\nu_n |n\rangle\langle n| + \sum_n t_n (|n\rangle\langle n+1| +|n+1\rangle\langle n|,
\end{equation}
where $n=1,2,\dots,21$ enumerates the resonators, $\nu_n=\nu=\nu_0-i\gamma$ is the resonance frequency of the $n$th individual resonator and $t_n(=t_1 \textrm(or) t_2)$ is the coupling between nearest resonators. The band-structure of the system of Eq.~\eqref{eq:eq1} consists of two mini-bands placed at frequency intervals $\nu_0-t_1-t_2<\nu<\nu_0- |t_1-t_2|$ and $\nu_0+|t_1-t_2 |<\nu<\nu_0+t_1+t_2$ separated by a finite gap of width $2|t_1-t_2|$. In the presence of the defect resonator at $n_0=11$ a topologically protected localised defect mode is created in the middle of the band at $\nu_D=\nu_0$ \cite{22}. The mode is exponentially localised at the defect resonator. Its shape, in the limit of infinite many resonators, is \cite{22,29}
\begin{equation}\label{eq:eq2}
\psi_n^D \sim \left\{
\begin{array}{cl}
\frac{1}{\sqrt{\xi}} e^{-|n-n_0|/\xi}; & n \textrm{ odd}\\
0; & n \textrm{ even}
\end{array}
\right.
\end{equation}
where $\psi_n^D$ is the amplitude of the defect mode at the $n$th resonator and $\xi=1/\ln(t_1/t_2)$ is the so-called localisation length of the mode \cite{22,29,asb16}.
The staggering form of the defect state is a direct consequence of the chiral (sublattice) symmetry. The chiral symmetry provides a topological protection of the defect mode to structural imperfections and at the same time shields the position of the resonant mode in the middle of the band-gap for reasonable variations of $t_1$ and $t_2$ \cite{22,asb16}.

\begin{figure}
  \centerline{\includegraphics[trim={0.8cm 0.25cm 1.2cm 2cm},clip,width=\columnwidth]{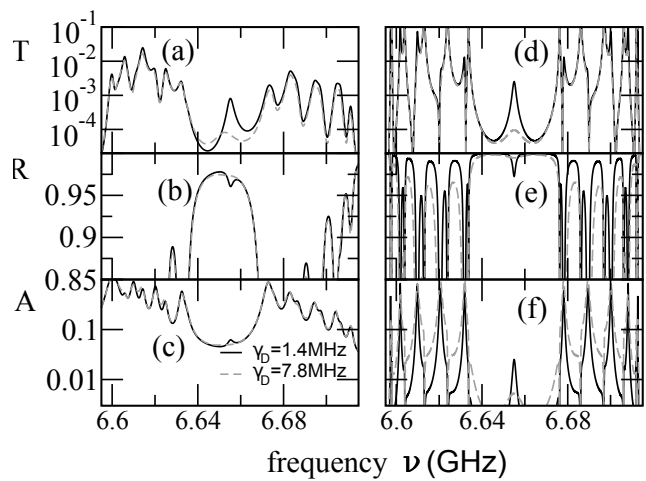}}
  \caption{\label{fig:fig2}
  {\bf Experimental and theoretical transmittances, reflectance, and absorption.}
  Microwave measurements for (a) the transmittance T; (b) the reflectance $R$; and (c) the absorption $A$ for the photonic structure of Fig.~\ref{fig:fig1}. We considered two different values of $\gamma_D=1.4$\,MHz (solid lines) and $\gamma_D=7.8$\,MHz (dashed lines) for the defect resonator. All other resonators have $\gamma=1.4$\,MHz. Numerical calculations for the (d) transmittance $T$; (e) Reflectance $R$ and (f) Absorption $A$. In these simulations we assumed that the resonators are lossless, i.e.\ $\gamma=0$, apart from the defect resonator which has the same amount of loss as at the left subfigures, i.e.\ $\gamma_D=1.4$\,MHz (solid lines) and $\gamma_D=7.8$\,MHz (dashed lines).
  }
\end{figure}

\begin{figure}
  \centerline{\includegraphics[trim={1.0cm 0.2cm 1.6cm 2.9cm},clip,width=\columnwidth]{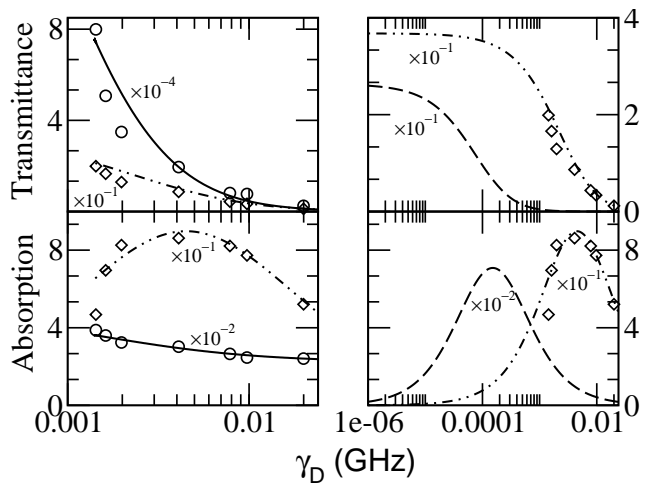}}
  \caption{\label{fig:fig3}
  {\bf Experimental and theoretical transmittance and absorption for single lossy resonator and CROWS as a function of loss strength $\gamma_D$.}
  (Left) The transmittance $T$ (top) and absorption $A$ (bottom) versus the linewidth $\gamma_D$ of the lossy defect resonator for the photonic structure (lines/circles) and for the stand-alone resonator (dashed-dotted lines/ diamonds). In the case of the photonic structure all other resonators have losses $\gamma=1.4$\,MHz. The lines indicate the numerical results for this set-up. (Right) Numerical results (lines) for the ideal photonic structure (dashed lines) with zero losses at all other resonators, i.e.\ $\gamma=0$. The symbols correspond to measurements of $T$ and $A$ for the case of a single lossy resonator.
}
\end{figure}

When the system is coupled to the antennas at the resonators $n_T=1$ (emitting antenna) and $n_R=13$ (receiving antenna), the localised defect mode becomes quasi-localised resonant mode at frequency $\nu_D \approx \nu_0$, with a large but finite lifetime $\tau$ (details see method section). One can estimate the inverse lifetime of such mode
\begin{equation}\label{eq:eq4}
\tau^{-1}\sim |w_T|^2 |\psi_1^D|^2+|w_R|^2 |\psi_{13}^D|^2\,,
\end{equation}
where $|\psi_1^D|^2,|\psi_{13}^D|^2$ are its $n=1$ and $n=13$ components, see Eq.~\eqref{eq:eq2}. $w_T$ and $w_R$ describe the coupling of the transmitting and receiving antenna, respectively. The measured transmittance $T=|S_{12}|^2$, reflectance $R=|S_{11}|^2$ and absorption $A=1-T-R$ versus frequencies $\nu$ of the CROW structure (with global $\gamma=1.4$\,MHz$=\gamma_D$) are shown in Figs.~\ref{fig:fig2}(a,b,c) (solid lines). Measurements of the widths of the mini-bands and of the gap allow us to extract the couplings $t_1=38$\,MHz, $t_2=21$\,MHz. We further see that the presence of the defect resonator results in a transmission peak at $\nu=\nu_D$ inside the photonic band gap. A fitting of the height of this peak (see method section Eq.~\eqref{eq:eq3}), for various $\gamma_D$ values, gives $w_T=10.915$\,MHz, $w_R=3.6875$\,MHz (see Fig.~\ref{fig:fig3}). The small peak in the absorption (solid line in Fig.~\ref{fig:fig2}(c)) is associated with the fact that all our resonators have a small ohmic component as we discussed above. In Fig.~\ref{fig:fig2}(a) we also report (dashed lines) our transport measurements for the case of a defect with additional losses, i.e.\ $\gamma_D=7.8$\,MHz. We find that even a small increase in the absorption coefficient $\gamma_D$ of the defect resonator can strongly suppress the phenomenon of resonant transmission, as illustrated in Fig.~\ref{fig:fig2}(a).

In Fig.~\ref{fig:fig2}(b) we show the reflectance of the CROW for $\gamma_D=1.4$\,MHz and $\gamma_D=7.8$\,MHz (solid and dashed lines, respectively). We find that the suppression in the resonant transmission is accompanied by an increase in the reflectivity of the CROW array. Moreover the incident light energy absorbed by the photonic structure is decreasing as $\gamma_D$ increases, see Fig.~\ref{fig:fig2}(c). In other words, our photonic structure becomes \emph{reflective} (not absorptive) as the losses of the defect resonator increases. This behavior is in distinct contrast to the case of a single (sacrificial) lossy resonator [see Fig.~\ref{fig:fig1}(e)] where the drop in transmittance is associated with an increase of absorption. These features are also observed in the simulations of an ideal CROW where all resonators have zero intrinsic losses $\gamma=0$, see Figs.~\ref{fig:fig2}(d,e,f).

An overview of the measurements (black circles) of $T(\nu_D)$, $A(\nu_D)$ and the corresponding numerical results (black solid lines) for the CROW array of Fig.~\ref{fig:fig1} versus the linewidth $\gamma_D$ of the defect lossy resonator are reported at the left column of Fig.~\ref{fig:fig3}. We find that an increase of $\gamma_D$ leads to a decrease of resonant transmittance $T$ and absorption $A$ of the photonic structure. This behaviour is contrasted with the measurements (diamonds) and numerical calculations (dashed lines) of a stand-alone lossy resonator. In the latter case we observe relatively large $T$ values $T\sim 10^{-1}$ as opposed to $T\sim 10^{-4}$ for the photonic structure, i.e.\ ultralow LT. For moderate $\gamma_D$-values the absorption of the stand-alone resonator reaches large values $A(\gamma_D=0.004\,\textrm{GHz})\approx 0.8$ corresponding to low LDT. In contrast, the CROW structure takes absorption values, which are at least one order of magnitude smaller (high LDT).
At the right column of Fig.~\ref{fig:fig3}, we report the numerical results for $T(\nu_D)$, $A(\nu_D)$ for an ideal CROW structure with $\gamma=0$ (dashed lines) and one lossy defect resonator with $\gamma_D$ placed in the middle of the array. Again, we compare these results to the theoretical/experimental (lines/diamonds) results for the stand-alone lossy resonator with linewidth $\gamma_D$. Both cases show the same qualitative behavior. However, the photonic structure shows a two-order lower LT (i.e.\ smaller $\gamma_D$-value for which the decay of transmittance occurs) as compared to a stand-alone resonator. At the same time the LDT of the photonic structure is at least two orders of magnitude higher than the corresponding one associated with the stand-alone resonator. Specifically the latter case acquires a maximum value of absorption $A\approx 0.8$ at $\gamma_D\approx 0.01$ as opposed to $A\approx 0.01$ acquired by the photonic structure. We point that the maximum in absorption for the photonic structure occurs at much lower values of $\gamma_D\sim 10^{-4}$ which in the case of a non-linear lossy mechanism correspond to rather small, and therefore harmless, incident radiation power/energy.

\begin{figure}
  \centerline{\includegraphics[trim={2.0cm 0.1cm 4.5cm 0.0cm},clip,width=\columnwidth]{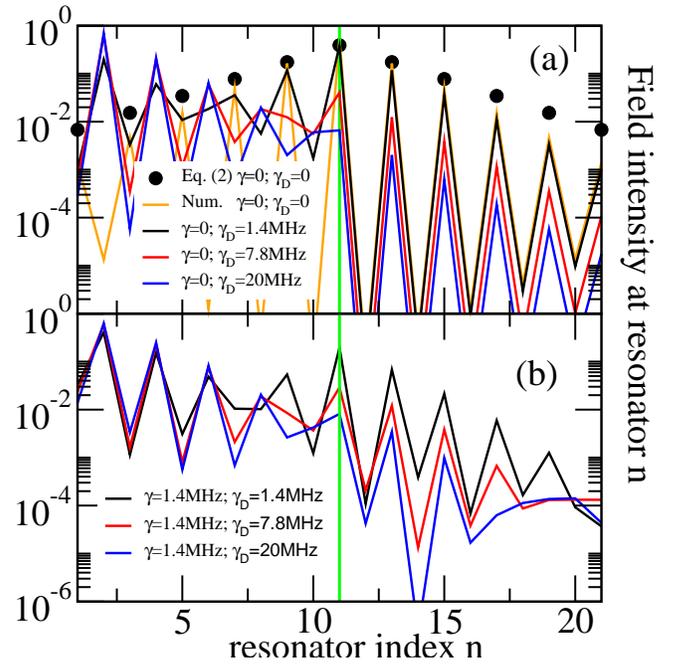}}
  \caption{\label{fig:fig4}
  {\bf Theoretical and experimental field intensities spatial profiles.}
  (a) Numerical results for an ideal CROW array consisting of lossless resonators with $\gamma=0$. Filled black circles correspond to the theoretical prediction Eq.~\eqref{eq:eq2} for the defect mode profile. Solid lines correspond to the simulations of the resonant defect mode profile for various loss-strengths $\gamma_D$. For symmetry reasons we assumed that the antennas are coupled to the first and last resonator. (b) Experimentally measured resonant mode profiles for various $\gamma_D$-values. The losses at all the resonators of the photonic structure are measured to be $\gamma=1.4$\,MHz.
  }
\end{figure}

The fragility of the resonant localised mode at moderate $\gamma_D$-values is further analysed in Fig.~\ref{fig:fig4}. In Fig.~\ref{fig:fig4}(a) we report the numerical resonant defect fields for a lossless resonator array (i.e.\ $\gamma=0$) and for various $\gamma_D$-values. For $\gamma_D=0$, a nice agreement between the numerics and Eq.~\eqref{eq:eq2} is observed, indicating that the coupling to the antennas does not affect the resonant mode profile. As the loss $\gamma_D$ at the defect resonator increases from zero, a gradual deviation from the profile of Eq.~\eqref{eq:eq2} occurs and eventually a suppression of the defect localized mode is observed.
Indeed, for $\gamma_D=20$\,MHz the exponential decay of the field intensity from the incident side of the waveguide [see blue line in Fig.~\ref{fig:fig4}(a)] indicates the destruction of the resonant localized mode [see filled circles in Fig.~\ref{fig:fig4}(a)].
\emph{Thus the lossy defect resonator is protected from damages induced by heat or electrical breakdown.} For the photonic limiter of Fig.~\ref{fig:fig1} it implies a huge increase in its dynamic range. The comparison with the experimental data [see Fig.~\ref{fig:fig4}(b)] associated with the CROW array of Fig.~\ref{fig:fig1} shows that the underlying mechanism which is responsible for the destruction of the resonant defect mode remains unaffected, even when small global losses are present ($\gamma=1.4$\,MHz).

The destruction of the resonant defect mode can be understood intuitively as a result of a competition between two mechanisms that control the dwell time of the photon in the resonant state. The first one is associated with the boundary losses due to the coupling of the photonic structure to the antennas. It results to a resonant linewidth $\Gamma_\textrm{edge}\sim \tau^{-1}$, see Eq.~\eqref{eq:eq4}. The other mechanism is associated with bulk losses and it leads to an additional broadening of the resonance linewidth. Using first order perturbation theory we get
\begin{equation}\label{eq:eq5}
\Gamma_\textrm{bulk}\approx \gamma_D |\psi_{11}|^2+\gamma \sum_{n\ne 11}|\psi_n|^2 =(\gamma_D-\gamma)/\xi+\gamma.
\end{equation}

For small values of $\gamma_D$ such that $\Gamma_\textrm{bulk}<\Gamma_\textrm{edge}$, the dwell time is determined by $\Gamma_\textrm{edge}$ and it is essentially constant. Thus the absorption of the photons that populate the resonant state increases, as they are trapped for relatively long time in the lossy CROW array [see the peak of the black line in Fig.~\ref{fig:fig2}(c)]. When $\Gamma_\textrm{bulk}\approx \Gamma_\textrm{edge}$, the dwell time itself begins to diminish, and the resonant mode is spoiled. For even larger values of $\gamma_D$ the photons do not dwell at all in the resonant state and reflection from the whole structure becomes the dominant mechanism. As a result, the absorption decreases to zero. The above argumentation applies equally well for the stand-alone defect and for the photonic structure. However, in the latter case the condition for the destruction of the resonant mode $\Gamma_\textrm{bulk}\approx\Gamma_\textrm{edge}$ is achieved for \emph{exponentially} smaller values of $\gamma_D$. It is exactly this effect that our proposal is harvesting in order to increase the damaging threshold (and consequently the dynamic range) of the photonic waveguide limiter.

\section*{Discussion}
We have investigated experimentally the transport properties of a coupled microwave resonator array with a lossy defect in the presence of chiral symmetry. We have shown that when the losses at the defect resonator are small, a topologically protected midgap defect mode provides high transmittance. When the losses increase, the defect mode and the associated resonant transmission are suppressed. We find that the incident energy is not absorbed - despite the fact that the defect losses are increased. Rather the incident radiation is reflected back to space. Such mechanism, coupled with non-linear losses at the defect resonator, can be harvested for the realisation of a new class of microwave waveguide limiters with a wide dynamical range.

\begin{acknowledgments}
The authors acknowledge partial support from ONR via grant N00014-16-1-2803 and from AFOSR (LRIR14RY14COR
and FA9550-14-1-0037). The stay of (T.K.) at LPMC-CNRS was supported by CNRS.
\end{acknowledgments}

\section{Methods}
\subsection{Antenna coupling to the system and linewidth extraction}
The transmission and reflection measurements are performed on the one hand-side with a kink antenna, which couples to the in-plane electric field, and on the other hand side with the loop antenna, which couples to the magnetic field which is oriented out-of-plane. The excited resonance of the resonators corresponds to the first resonance of the TE$_1$ mode, i.e.\ the magnetic field is out-of-plane and maximal in the center and the electric field is in-plane and maximal at the boundaries of the resonators. Thus the two antenna types are ideally suited to perform the experiment. To obtain the natural line width of the resonators, i.e.\ without the linewidth acquired by the antenna coupling, we measured the resonance linewidth as a function of the distance between antenna and resonator and extrapolated the natural linewidth.

\subsection{Model to describe the coupling to environment}
The photonic structure is coupled to the antennas at the  $n_T=1$ resonator and at the $n_R=13$ resonator by a transmitting and a receiving antenna, respectively (see Fig.~\ref{fig:fig1}). We are modelling the two antennas by a one-dimensional semi-infinite tight-binding lattice with coupling constant $t_L=(t_1+t_2 )/2$ and on-site energies $\nu_L=\nu_0$. The associated scattering matrix can be written in the following form \cite{30}
\begin{eqnarray}\label{eq:eq3}
\hat{S}&=&-\hat{1}+\frac{2i \sin k}{t_L} W^T \frac{1}{H_{eff}-\nu} W;   \\
H_{eff}&=&H_P+\frac{e^{ik}}{t_L} WW^T,
\end{eqnarray}
where $\hat{1}$ is the $2\times 2$ identity matrix, $W_{nm}=w_T\delta_{n,n_T} \delta_{m,1} +w_R \delta_{n,n_R} \delta_{m,2}$ is a $N\times 2$ matrix that describes the coupling between the photonic structure and the antennas, $\nu=\nu_L+2t_L\cos k$ is the frequency of propagating waves at the antennas and $k$ is their associated wavevector. Finally $w_{T}$ and $w_{R}$ are the coupling constant of the transmitting and the receiving antennas, respectively. Due to the coupling to the antennas, the localised defect mode becomes quasi-localised resonant mode at frequency $\nu_D \approx \nu_0$, with a large but finite lifetime $\tau$. One can estimate (to first order in perturbation theory) the inverse lifetime of such mode
\begin{equation}\label{eq:eq4}
\tau^{-1}\sim \left\langle \psi^D\left|\frac{e^{ik}}{t_L} WW^T\right|\psi^D\right\rangle
=|w_T|^2 |\psi_1^D|^2+|w_R|^2 |\psi_{13}^D|^2\,,
\end{equation}
where $|\psi_1^D|^2,|\psi_{13}^D|^2$ are its $n=1$ and $n=13$ components, see Eq.~\eqref{eq:eq2}.

\end{document}